\documentclass[10pt]{iopart}
\usepackage{graphicx}
\usepackage{latexsym}

\usepackage{iopams}
\def\sn1987a{SN \,1987A\,}
\def\etacar{$\eta$-Carinae~}
\def\duesnr{SN\, 2009ig~}
\def\tresnr{SN\, 2009bb~}
\def\ca{\rm{CA}\,{\small\rmfamily II}\,}
\def\velu{\rm\,{kms^{-1}}}
\def\sn1987a{SN \,1987A\,}
\def\apj{ApJ\,  }

\def\nat{Nature\,  }

\begin{document}
\title
{
The relativistic  three dimensional evolution of  SN 1987A
}
\vskip  1cm
\author     {L. Zaninetti}
\address    {Dipartimento  di Fisica,
 via P.Giuria 1,\\ I-10125 Turin,Italy }
\ead {zaninetti@ph.unito.it}

\begin {abstract}
The high velocities observed in supernovae require
a relativistic treatment for the equation of motion
in the presence of gradients  in the density of the interstellar medium.
The adopted theory  is that of the thin layer approximation.
The chosen medium is auto-gravitating  with respect to an equatorial plane.
The  differential equation which governs  the relativistic
conservation of momentum  is solved in numerically
and by recursion. The asymmetric field of relativistic velocities
as well the time dilation are plotted at the age of 1 yr for
SN 1987A.
\end{abstract}
\vspace{2pc}

\maketitle

\section{Introduction}

The expansion velocities in supernovae (SN) are quite high
and, as an example, a  time series of eight  spectra in \duesnr
reports that  the velocity at the  \ca line,
decreases from 32000 $\velu$ to 21500$\velu$,
in  12 day,
see Fig. 9 in \cite{Marion2013}.
Another example is given
by  \tresnr  in which the velocity
of expansion has  been evaluated
to be
$\approx$ 255000 $\velu$, see \cite{Soderberg2010}.

We briefly recall that the corrections in special relativity (SR)
for stable atomic clocks in satellites
of the
Global Positioning System (GPS) are applied to satellites which are moving
at a velocity of $\approx 3.87 \velu$.
The problem of the aspherical SN, such as \sn1987a,
is to find an acceptable  model which can reproduce
the observed complex morphology of the  aspherical \sn1987a
and this was done in a classical framework by
\cite{Zaninetti2013c}.
In this paper we shall discuss a relativistic treatment
of the thin layer approximation in the presence of an auto-gravitating medium.

\section{Relativistic conservation of momentum}

The chosen auto-gravitating profile is
\begin{equation}
n(R,\theta) = n_0 sech^2 (\frac{R \sin (\theta) }{2\,h})
\quad ,
\label{sech2rtheta}
\end{equation}
where
$n_0$ is the density in the equatorial plane ($\theta=0$),
$R$ is the radius of the advancing shell,
$\theta$  is
the latitude angle  ($\theta=0$ at the equator and $\theta=\pm 90$
at the two poles)
and $h$ is a parameter which characterizes the gradient.
The chosen symmetry imposes that the  motion is independent   of the azimuthal angle in spherical coordinates but depends only on the latitude angle and the time.
The classical  conservation of momentum
in the presence of an auto-gravitating medium
was treated in  \cite{Zaninetti2013c} and therefore
we will not duplicate the results already obtained.
The relativistic  conservation of momentum,
see \cite{French1968,Zhang1997,Guery2010},
is formulated as
\begin{equation}
M(R_0;b) \gamma_0 \beta_0 = M(R;b) \gamma \beta
\quad ,
\label{momentumrelativistic}
\end{equation}
with
\begin{equation}
\gamma_0 = \frac{1} {
\sqrt{1-\beta_0^2}
}
\quad ; \qquad
\gamma = \frac{1} {
\sqrt{1-\beta^2}
}
\quad ,
\end {equation}
and
\begin{equation}
\beta_0 =\frac{v_0}{c}
\quad ; \qquad
\beta =\frac{v}{c}
\quad  ,
\end{equation}
$c$  being  the velocity of light,
here $M(R_0;b)$ is a first  mass between 0 and $R_0$ and
$M(R;b)$ is a second  mass between 0 and $R$.
We know already that $M(R;b) = (I_m(R))^{1/p}$
where the integral $I_m(R)$ has been defined in eq. (15) of
\cite{Zaninetti2013c} and $p$ is a parameter to be found.
The fundamental Eq. (\ref{momentumrelativistic})  can be  first
solved for $\beta^2$
\begin{equation}
\beta^2 = \frac{N}{D}
\quad ,
\end{equation}
where
\begin{eqnarray}
N ={{\it R_0}}^{6\,{p}^{-1}}{{\it \beta_0}}^{2}
\quad ,
\nonumber
\end{eqnarray}
\begin{eqnarray}
D= -  ( -1  ) ^{2\,{p}^{-1}}  ( -C{{\it R0}}^{3}{S}^{3}+8\,C
\ln   ( 1+B  ) R{h}^{2}S
\nonumber \\
-8\,C\ln   ( 1+A  ) {\it R0}
\,{h}^{2}S+4\,C{{\it R0}}^{2}h{S}^{2}-4\,C{R}^{2}h{S}^{2}
\nonumber \\
-B{{\it R0}}^
{3}{S}^{3}-A{{\it R0}}^{3}{S}^{3}+8\,CP  ( 2,-B  ) {h}^{3}-8
\,CP  ( 2,-A  ) {h}^{3}
\nonumber \\
+8\,B\ln   ( 1+B  ) R{h}^{2}S
-8\,B\ln   ( 1+A  ) {\it R0}\,{h}^{2}S-4\,B{R}^{2}h{S}^{2}
\nonumber \\
+8
\,A\ln   ( 1+B  ) R{h}^{2}S-8\,A\ln   ( 1+A  ) {\it
R0}\,{h}^{2}S+4\,A{{\it R0}}^{2}h{S}^{2}
\nonumber \\
-{{\it R0}}^{3}{S}^{3}+8\,BP
  ( 2,-B  ) {h}^{3}-8\,BP  ( 2,-A  ) {h}^{3}+8\,AP
  ( 2,-B  ) {h}^{3}
\nonumber \\
  -8\,AP  ( 2,-A  ) {h}^{3}+8\,{h}^
{2}R\ln   ( 1+B  ) S-8\,{h}^{2}{\it R0}\,\ln   ( 1+A
  ) S
  \nonumber \\
  +8\,{h}^{3}P  ( 2,-B  ) -8\,{h}^{3}P  ( 2,-A
  )   ) ^{2\,{p}^{-1}}  ( 1+B  ) ^{-2\,{p}^{-1}}{S}
^{-6\,{p}^{-1}}  ( 1+A  ) ^{-2\,{p}^{-1}}{{\it \beta_0}}^{2}
\nonumber \\
+
  ( -1  ) ^{2\,{p}^{-1}}  ( -C{{\it R_0}}^{3}{S}^{3}+8\,C
\ln   ( 1+B  ) R{h}^{2}S-8\,C\ln   ( 1+A  ) {\it R_0}
\,{h}^{2}S
\nonumber \\
+4\,C{{\it R_0}}^{2}h{S}^{2}-4\,C{R}^{2}h{S}^{2}-B{{\it R_0}}^
{3}{S}^{3}-A{{\it R_0}}^{3}{S}^{3}+8\,CP  ( 2,-B  ) {h}^{3}
\nonumber \\
-8
\,CP  ( 2,-A  ) {h}^{3}+8\,B\ln   ( 1+B  ) R{h}^{2}S
-8\,B\ln   ( 1+A  ) {\it R_0}\,{h}^{2}S-4\,B{R}^{2}h{S}^{2}
\nonumber \\
+8
\,A\ln   ( 1+B  ) R{h}^{2}S-8\,A\ln   ( 1+A  ) {\it
R_0}\,{h}^{2}S+4\,A{{\it R_0}}^{2}h{S}^{2}-{{\it R_0}}^{3}{S}^{3}
\nonumber \\
+8\,BP
  ( 2,-B  ) {h}^{3}-8\,BP  ( 2,-A  ) {h}^{3}+8\,AP
  ( 2,-B  ) {h}^{3}-8\,AP  ( 2,-A  ) {h}^{3}
 \nonumber \\
  +8\,{h}^
{2}R\ln   ( 1+B  ) S-8\,{h}^{2}{\it R_0}\,\ln   ( 1+A
  ) S+8\,{h}^{3}P  ( 2,-B  )
  \nonumber \\
  -8\,{h}^{3}P  ( 2,-A
  )   ) ^{2\,{p}^{-1}}  ( 1+B  ) ^{-2\,{p}^{-1}}{S}
^{-6\,{p}^{-1}}  ( 1+A  ) ^{-2\,{p}^{-1}}+{{\it R_0}}^{6\,{p}^
{-1}}{{\it \beta_0}}^{2}
\quad ,
\nonumber
\end{eqnarray}
with
\begin{eqnarray}
A= &     {{\rm e}^{{\frac {R_{{0}}\sin \left( \theta \right) }{h}}}}
\nonumber \\
B= & {{\rm e}^{{\frac {R\sin \left( \theta \right) }{h}}}}
\nonumber \\
C= & {{\rm e}^{{\frac {\sin \left( \theta \right)  \left( R_{{0}}+R
 \right) }{h}}}}
 \nonumber \\
 S= & \sin \left( \theta \right)
 \nonumber
 \quad .
 \end{eqnarray}
and  $P$  the polylog operator, which is
defined by
\begin{equation}
polylog(a,z) = \sum
_{{n=1}}^{\infty}\frac{z^{n}}{n^{a}}
\quad .
\end{equation}
The value of $\beta$ is
\begin{equation}
\beta = \sqrt{\frac{N}{D}}
\quad ,
\end{equation}
or
\begin{equation}
\frac{dR}{dt} = c \sqrt{\frac{N}{D}}
\label{eqndiff}
\quad .
\end{equation}
This first order  differential equation can be solved with
the Runge--Kutta method, see
FORTRAN SUBROUTINE  RK4 in \cite{press}.
Another approach
separates the variables
\begin{equation}
\int_{R_0}^R
\frac{1}{\sqrt{\frac{N}{D}}}
\,dR
= c (t-t_0)
\quad .
\label{nlequation}
\end{equation}
The previous integral does not have an analytical solution
and we treat the previous result as a non-linear equation to be
solved with the FORTRAN SUBROUTINE  ZRIDDR in \cite{press}.
The presence of an analytical  expression for $\beta$
as given  by Eq. (\ref{eqndiff})  allows setting up the
recursive solution
\begin{eqnarray}
R_{n+1} =   &  R_n + V_n(R_0,R_n,\beta_0,h) \Delta t    \nonumber  \\
V_{n+1} =   &  V_n(R_0,R_{n+1},\beta_0,h)
\quad  ,
\label{recursiverel}
\end{eqnarray}
where  $R_n$, $V_n$, $\Delta t$ are the temporary  radius,
the relativistic velocity,
and the interval of time, respectively.
An interesting application of  SR  is the time delay:
given an interval  of time, $\Delta t$,
in the laboratory  frame the interval of time,
$\Delta t^{\prime} $,
in a frame that that is  moving with
velocity  $v$ in the $x$-direction is
\begin{equation}
\Delta t^{\prime} = \frac{\Delta t}{\sqrt{1-\frac{v^2}{c^2}}}
\quad  .
\end{equation}
We can therefore introduce the following ratio
\begin{equation}
D = \frac{\Delta t } {\Delta t^{\prime}}
\quad ,
\label{dilation}
\end{equation}
which measures the time dilation, and lies between 0 and 1.

\section{Astrophysical application}

We numerically solved the
non-linear equation, Eq. (\ref{nlequation})
even if the same results  can be obtained
by solving the  differential equation
(\ref{eqndiff})
or implementing   the
recursive  relationship
as given by
Eq. (\ref{recursiverel}), see
Table \ref{datafitrel1987a} for the
adopted data.
\begin{table}
\caption
{
The numerical values of the parameters
of the relativistic simulation  for
\sn1987a
}
\label{datafitrel1987a}
 \[
 \begin{array}{lcc}
 \hline
 \hline
\mbox{Quantity}     & \mbox{Unit}         &   \mbox{value} \\
R_0          & \mbox{pc}           &   0.011 \\
\dot {R}_0   & \mbox{km s}^{-1} &   30000 \\
p            & \mbox{number}       &   4       \\
h            & \mbox{pc}           &   0.01     \\
t_{0}        & \mbox{yr}          &   0.00022  \\
t            & \mbox{yr}         &   23       \\
\noalign{\smallskip}
 \hline
 \hline
 \end{array}
 \] 
 \end {table}

\begin{figure}
  \begin{center}
\includegraphics[width=10cm]{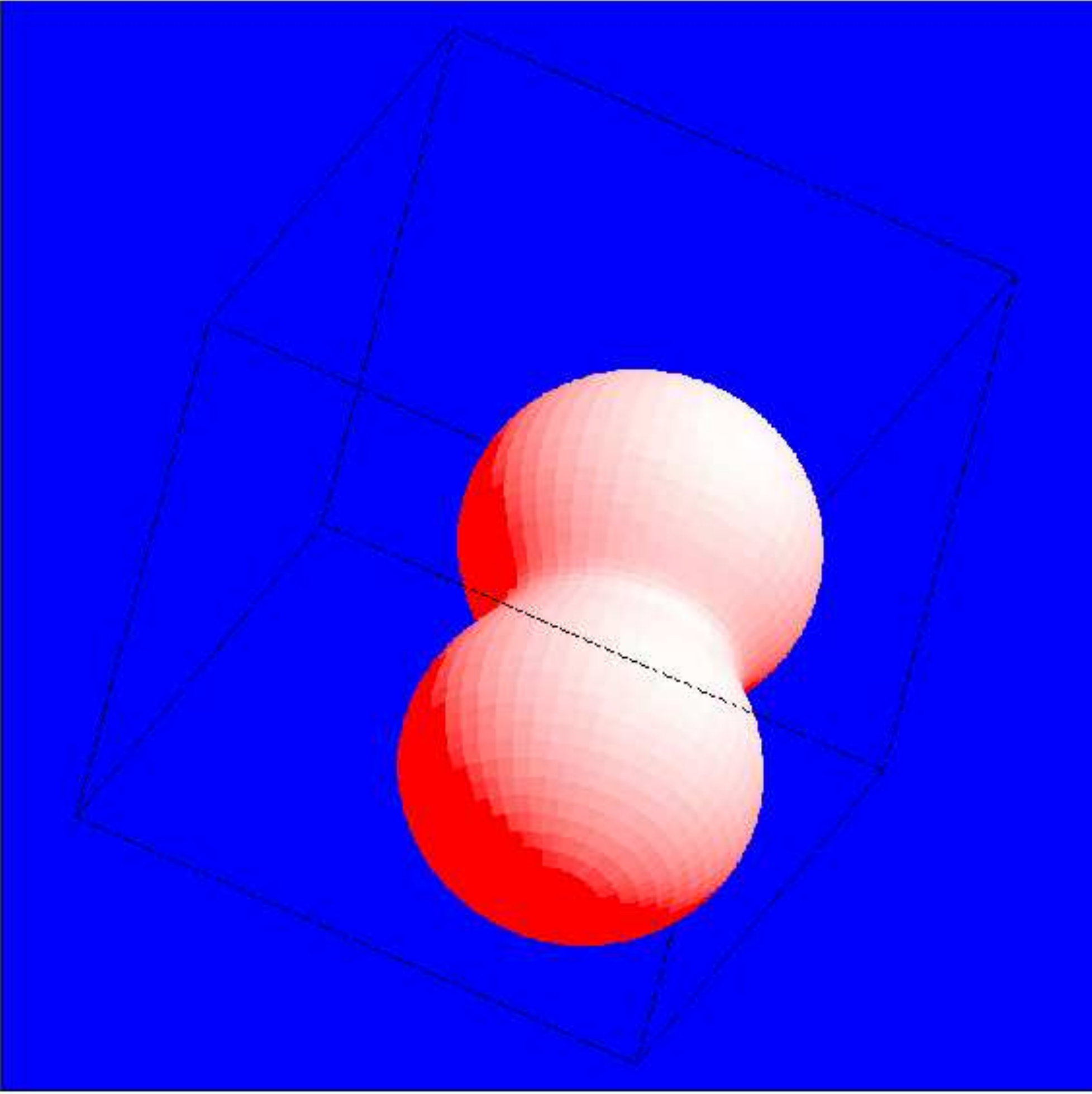}
  \end {center}
\caption {
 Continuous  three-dimensional surface of \sn1987a
 after 23 yr: the three Eulerian angles characterizing the point of view are
     $ \Phi   $=105$^{\circ }$,
     $ \Theta $=55 $^{\circ }$
and  $ \Psi   $=-165 $^{\circ }$.
Physical parameters as in Table
\ref{datafitrel1987a}.
          }%
    \label{rel_sn1987a_faces_rotated}
    \end{figure}

\begin{figure}
\begin{center}
\includegraphics[width=6cm]{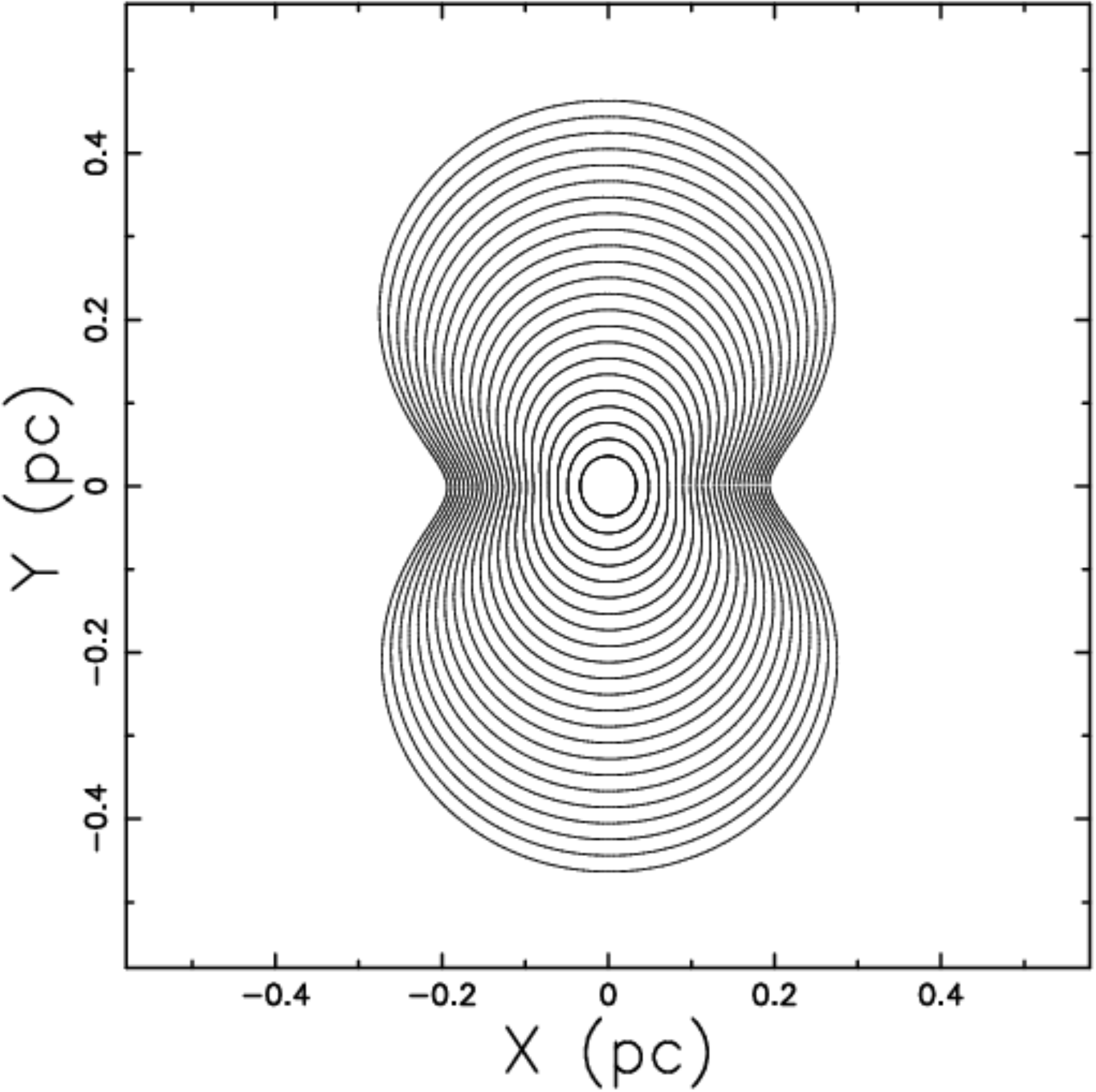}
\end{center}
\caption
{
Sections  of \sn1987a  in the {\it X-Z}  plane
at time steps of 1yr.
Physical parameters as in Table \ref{datafitrel1987a}.
This is a non-rotated image
and the  three Euler angles
characterizing the   orientation
are $ \Phi $=180$^{\circ }$,
$ \Theta     $=90 $^{\circ }$
and   $ \Psi $=0  $^{\circ }$.
}
\label{rel_sezioni_1987a}%
    \end{figure}

\begin{figure}
\begin{center}
\includegraphics[width=6cm]{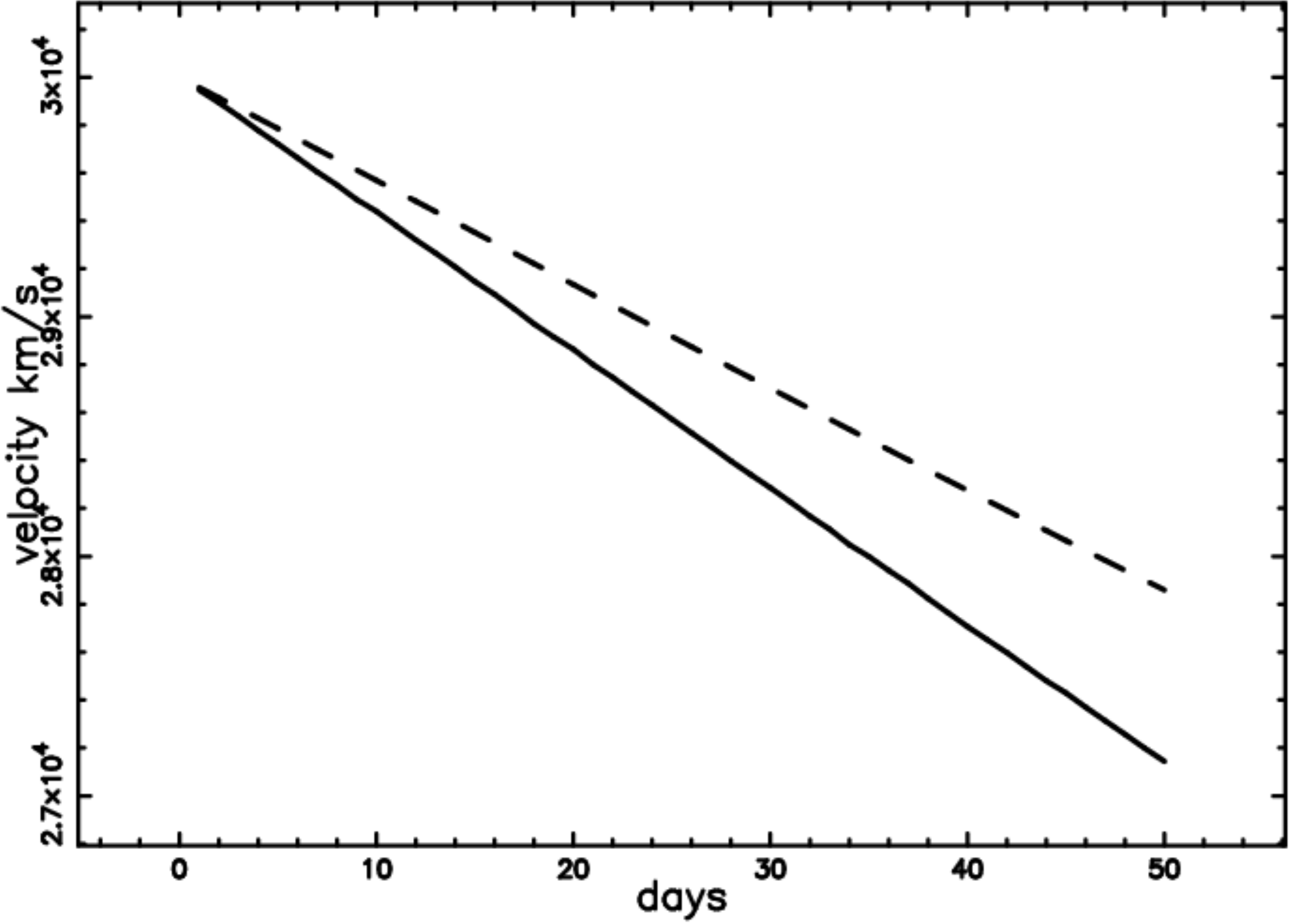}
\end{center}
\caption
{
Velocity  in the  equatorial  direction (full line)
and       in the  polar       direction (dashed line)
in the first 50 days.
}
\label{duevel_sn1987a}%
    \end{figure}

\begin{figure}
  \begin{center}
\includegraphics[width=6cm]{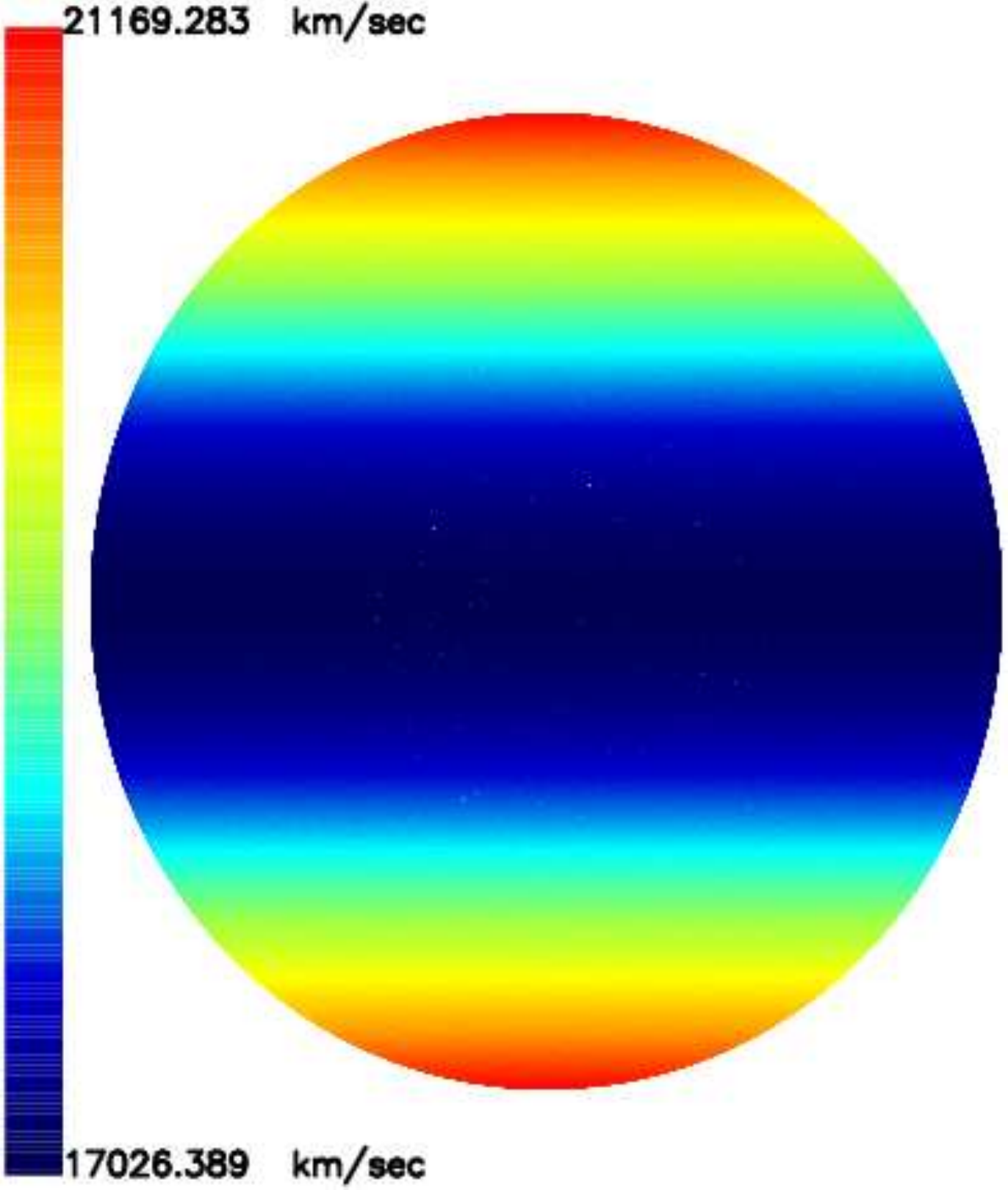}
  \end {center}
\caption
{
Map of the relativistic  velocity
as a function of the latitude
for  \sn1987a at the age  1 yr.
}%
\label{sn1987a_v_field_rel}
\end{figure}

\begin{figure}
  \begin{center}
\includegraphics[width=6cm]{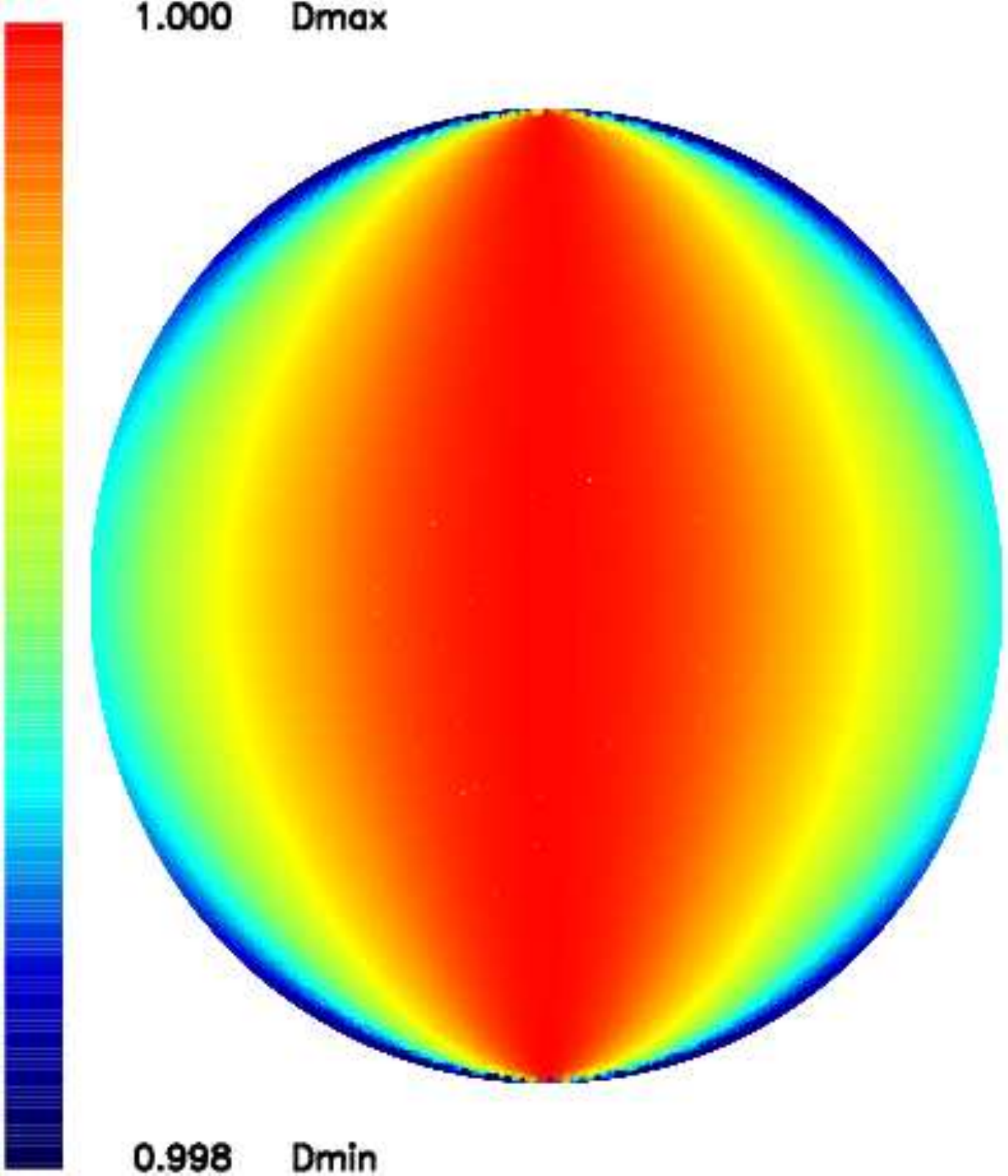}
  \end {center}
\caption
{
Map  of the relativistic time dilation
$D$, see Eq. (\ref{dilation}),
for velocity of  \sn1987a   in the direction perpendicular
to the observer at the age  1 yr at time intervals
of 1 s in the laboratory frame.
}
\label{timedilation}
\end{figure}

The complicated  structure  of \sn1987a
is due to the great variety of shapes
obtained when the point
of view of the observer
changes.
One way to parametrize the point of view
of the observer
is  the introduction of
the Euler angles $(\Phi, \Theta, \Psi)$,
as an example, Fig. \ref{rel_sn1987a_faces_rotated}
shows the 3D advancing shell
after  23 years.
In order to avoid complicated changes
of framework for the field of velocity
we limit ourselves to the non-rotated image.
This choice  is already widely used by 
astronomers in order  to reduce the data
of \etacar,  see Fig. 4 in \cite{Smith2006}.
The  progressive  increase of the asymmetry
is clearly outlined in Fig.
\ref{rel_sezioni_1987a}, in which sections
of the expansion are drawn at time steps of 1 yr.
The difference in velocity between
the polar direction and equatorial direction
are oulined in  Fig. \ref{duevel_sn1987a}.
The relativistic field  of velocity in the various
points of \sn1987a
after  1 yr was shown
in Fig. \ref{sn1987a_v_field_rel}.
The relativistic time dilation  is mapped in
Fig. \ref{timedilation} where
the velocity  of expansion  perpendicular  to
the observer ($x$-direction) is considered.

\section{Conclusions}

We have covered the evolution of a SN in an auto-gravitating medium
in a relativistic framework. The initial shape is represented
by a sphere of radius  $R_0=0.011$\ pc.
After 1 yr, the asymmetry  between the radius
in the equatorial plane
and the radius in the polar direction is
well defined
and Fig. \ref{sn1987a_v_field_rel}
summarizes both the asymmetrical
shape and the anisotropic field  of velocity.
The time dilation at  1 yr as represented
by the parameter $D$ varies
between  a minimum of  0.9975   and a maximum 
of 1.

\noindent
{\bf REFERENCES}


\begin{thebibliography}{1}
\expandafter\ifx\csname url\endcsname\relax
  \def\url#1{{\tt #1}}\fi
\expandafter\ifx\csname urlprefix\endcsname\relax\def\urlprefix{URL }\fi
\providecommand{\eprint}[2][]{\url{#2}}

\bibitem{Marion2013}
{Marion} G~H, {Vinko} J and {Wheeler} J~C 2013 {High-velocity Line Forming
  Regions in the Type Ia Supernova 2009ig} {\em \apj\/} {\bf 777} 40

\bibitem{Soderberg2010}
{Soderberg} A~M, {Chakraborti} S and {Pignata} G 2010 {A relativistic type Ibc
  supernova without a detected {$\gamma$}-ray burst} {\em \nat\/} {\bf 463},
  513 (\textit{Preprint} \eprint{0908.2817})

\bibitem{Zaninetti2013c}
{Zaninetti} L 2013 Three dimensional evolution of SN 1987a in a
  self-gravitating disk {\em International Journal of Astronomy and
  Astrophysics\/} {\bf 3}, 93

\bibitem{French1968}
{{French}, AP} 1968 {\em {Special Relativity}\/} (New~York: {CRC})

\bibitem{Zhang1997}
{Zhang} Y 1997 {\em Special Relativity and Its Experimental Foundations\/}
  (Singapore: World Scientific)

\bibitem{Guery2010}
Gu{\'e}ry-Odelin D and Lahaye T 2010 {\em Classical Mechanics Illustrated by
  Modern Physics: 42 Problems with Solutions\/} (London: Imperial College
  Press)

\bibitem{press}
{Press} W~H, {Teukolsky} S~A, {Vetterling} W~T and {Flannery} B~P 1992 {\em
  {Numerical Recipes in FORTRAN. The Art of Scientific Computing}\/}
  (Cambridge: Cambridge University Press)

\bibitem{Smith2006}
{Smith} N 2006 {The Structure of the Homunculus. I. Shape and Latitude
  Dependence from H2 and Fe II Velocity Maps of eta Carinae} {\em \apj\/} {\bf
  644}, 1151 (\textit{Preprint} \eprint{arXiv:astro-ph/0602464})

\end{thebibliography}
\providecommand{\newblock}{}

\end{document}